# Concept-Based Generic Programming in C++

Bjarne Stroustrup (www.stroustrup.com)
Columbia University

## Abstract

We present programming techniques to illustrate the facilities and principles of C++ generic programming using concepts. Concepts are C++'s way to express constraints on generic code. As an initial example, we provide a simple type system that eliminates narrowing conversions and provides range checking without unnecessary notational or run-time overheads.

Concepts are used throughout to provide user-defined extensions to the type system. The aim is to show their utility and the fundamental ideas behind them, rather than to provide a detailed or complete explanation of C++'s language support for generic programming or the extensive support provided by the standard library.

Generic programming is an integral part of C++, rather than an isolated sub-language. In particular, key facilities support general programming as well as generic programming (e.g., uniform notation for types, lambdas, variadic templates, and C++26 static reflection).

Finally, we give design rationales and origins for key parts of the concept design, including use patterns, the relationship to Object-Oriented Programming, value arguments, notation, concept type-matching, and definition checking.

## 1. Introduction

Generic programming is a style of programming where commonalities among types are abstracted into requirements known as concepts and typically used in code to constrain arguments to generic types and functions (often called algorithms). For example:

> **void sort(Sortable_range auto& r);**

This declares an algorithm **sort** that can take an argument of any type that meets the requirements of the concept **Sortable_range**.

Alex Stepanov, the father of modern C++ generic programming, expressed his goal for generic programming like this:

> *"Aim: The most general, most efficient, most flexible representation of concepts."*

For much of the code where generic programming is relevant, such as foundational code and code with real-time constraints, both flexibility and performance are major concerns. To be viable, the design and programming techniques must be widely useful as well as affordable in compile time, run time, and memory consumption in their intended domains. They must obey the zero-overhead principle [BS'94].

This paper presents the general idea of generic programming as it is supported by contemporary C++, addressing three design requirements outlined in 1994 [BS'94]:

- **Generality**: "Must be able to express more than I can imagine."
- **Uncompromised efficiency**: Generic code must not impose run-time overheads compared to roughly equivalent lower-level code. E.g., using a generic, strongly-typed vector is as efficient as equivalent uses of a C array.
- **Statically type-safe interfaces**: The type system must be flexible enough to allow compile-time checking of most aspects of interfaces that do not depend on run-time values.

Further requirements include:

- **Affordable**: Doesn't require expensive computers or slow compilers.
- **Teachable**: "if we require PhDs from MIT, we have failed." – Kristen Nygaard, the father of object-oriented programming,

Naturally, C++'s support for generic programming is not perfect, but it is in many ways better than alternatives. In particular, concepts can handle combinations of types (§5) and combinations of types and values (§7.5). Here, language facilities and fundamental techniques are presented through examples accompanied by rationales. The approach is like teaching a natural language through useful examples, rather than first explaining the complete grammar and vocabulary of that language. All technical details can be found in the ISO C++ standard [C++23], but that document is most certainly not a tutorial. Finally (§7), some design decisions for concepts are discussed.

## 2. Arithmetic conversions

In generic programming, implementations are chosen to match the arguments used. For example:

```
void print(auto arg);     // a generic function

void test()
{
    print (1);            // arg is an int
    print ("one");        // arg is a C-style string
}
```

This leaves the possibility of type errors to type conversions. Explicit conversions are unavoidable, but very rare in well-designed code, but implicit conversions are hard to avoid and can lead to loss of information with functions taking fixed static types. For example:

```
void print(int);
void print(char);

void f(unsigned x, int y)
{
    print(x);     // can x fit into an int?
    print(y);     // can y fit into a char?
}
```

Implicit narrowing conversions among arithmetic types are convenient and demanded by many users, but also a nasty source of errors. For C++, they arose in the early days before C had explicit type conversions (casts). Using generic programming, we can build a set of arithmetic types (of various sizes) that allows only conversions which don't change values (non-narrowing conversions). The problems with conversions that change values in surprising ways come in a few flavors:

- An integer converted to a type with too few bits to represent its value (e.g., **short x = 1'000'000;** assuming that a **short** is 16 bits and **float y = 0xFFFF'FFFFu;** assuming a **float** is 32 bits).
- An unsigned integer with a representation interpreted as a (large) negative integer after conversion to a signed value (e.g., **short x = 0b1000'0000'0000'0000u;**).
- A negative integer interpreted as a (large) positive value after conversion to an unsigned (e.g., **unsigned x = -2;**).
- A floating-point value with a decimal part truncated to its integer part when converted to an integer (e.g., **int x = 7.8;**).

In general, dealing with this requires looking at the values to be converted. However, in real code hardly any assignments lead to narrowing conversions. So, for performance reasons, we want to do run-time checks only where a narrowing conversion is possible.

Modern compilers warn against most simple and obvious examples like the ones above, but not all examples are simple and obvious. We have, of course, ways of explicitly checking for narrowing. However, to benefit from explicit testing we have to decide to test and then do testing consistently.

Let's start by dealing with only arithmetic types, that is, signed integers, unsigned integers, and floating-point numbers. We define a concept for that:

```
template<typename T>
concept Num = integral<T> || floating_point<T>;
```

A **concept** is a predicate (a function returning a **bool**) that can be evaluated at compile time and can take types as arguments. The **integral** and **floating_point** concepts are part of the standard library. For historical reasons, parameterization with types is indicated by angle brackets so in **integral<T>**, **<T>** indicates the type argument **T**. So, **integral<T>** means "is **T** an **integral** type?"

Later (§5), we explain how such predicates can be defined. If we need to, we can extend **Num** to also accept user-defined types, such as **Bigint** and **complex** (§6.4).

In C++, the **char** and **bool** types are deemed arithmetic. If we wanted to, we could deal with the problems that can arise from that also.

## 2.1. Detecting narrowing

It is crucial to be able to test whether assignment between a pair of types can lead to narrowing, so we define a **concept** for that:

```
template<typename T, typename U>
concept Can_narrow_to =      // can converting a T to a U cause loss of information?
  ! same_as<T, U>                    // different types
  && Num<T> && Num<U>                // both Nums
  && (
```

```
            (floating_point<T> && integral<U>)         // might dismiss a fractional part
            || (numeric_limits<T>::digits > numeric_limits<U>::digits)       // might truncate
            || (signed_integral<T>!=signed_integral<U> && sizeof(T)==sizeof(U)) // might change sign
        );
```

**Can_narrow_to<T,U>** is **true** if there exists a **T** value that loses information or change meaning if assigned to a **U**.

The **(numeric_limits<T>::digits > numeric_limits<U>::digits)** deals with the case where **T** is an integer type and **U** is an integer type with fewer bits (e.g. **short x = 234'567;**) and the case where **T** is a floating-point type with a mantissa is too small to hold the digits from a **U (**e.g., **Bfloat16_t x = 300;** [Intel]**)**. Floating-point types where that is possible are now common in graphics and AI/LLM code.

The **signed_integral<T>!=signed_integral<U>) && (sizeof(T)==sizeof(U)** deals with the fact that signed and unsigned integers of the same size have different value ranges.

The complexity of this directly reflects the complexity of modern hardware.

Finally, we need a function for testing values:

```
        template<Num U, Num T>
        constexpr bool will_narrow(U u)         // converting t to a U will cause loss of information
        {
            if constexpr (!Can_narrow_to<T, U>)
                return false;
            if constexpr (signed_integral<T> && unsigned_integral<U>)
                if (numeric_limits<T>::max() < t)        // too large positive?
                    return true;
            if constexpr (unsigned_integral<T> && signed_integral<U>)
                if (t<0)                                 // negative?
                    return true;
            T t = u;                                     // potentially narrows
            return (t != u);                             // narrows?
        }
```

First, we test what can be done at compile-time (the **if constexpr** tests). This makes trivial cases, such as assignments to the same or a larger type, free. Only if a change of value is possible do we use a run-time test. That means that we don't have to think too hard to decide whether to apply **will_narrow** and won't risk bugs to be introduced by changes to the types used during maintenance.

We can now define a function that tests for narrowing and decides what to do when it happens:

```
        class Bad_value {};

        template<Num T, Num U>
        constexpr T convert_to(T t)
        {
            if (will_narrow<U>(t))
                throw Bad_value{};
```

```
        return U(t);    // we only get to here if the cast doesn't narrow
}
```

I use exceptions to report errors because narrowing is very rare and explicitly checking every potential narrowing would be tedious, costly, and likely not done consistently (aka "error code hell"). Attempts to achieve complete error handling through explicit checks everywhere usually fail. In particular, the rules of implicit conversions among numeric types (mostly inherited from ancient C) are tricky and often only partially understood by developers. The points where they might occur are not easily spotted by a programmer. Also, much of the worry about exceptions is misinformed and mistaken (e.g., see [BS'22, KE'24]).

Now we can write:

```
void test(int si, char ch, unsigned ui, double d)
{
    auto x0 = convert_to<int>(si);        // redundant
    auto x1 = convert_to<int>(ui);        // ui could be too large
    auto x2 = convert_to<char>(si);       // si could be too large
    auto x3 = convert_to<int>(ch);        // redundant
    auto x4 = convert_to<unsigned>(ch);   // redundant or ch could be negative
    auto x5 = convert_to<unsigned>(si);   // si could be negative
    auto x6 = convert_to<double>(si);     // redundant if sizeof(int)<sizeof(double)
    auto x7 = convert_to<int>(d);         // could truncate (e.g., if d is 7.8)
}
```

The keyword **auto** is used to represent the type of the initializer saving us from repeating that type's name.

## 2.2. Making checking implicit

Those **convert_to** calls are verbose and tedious. They are unlikely to be used consistently, so let's provide a way to make them implicit.

```
template<Num T>
class Number {            // a Number<T> is a T that doesn't suffer narrowing conversions
    T val;
public:
    template<Num U>
    constexpr Number(const U u) : val{convert_to<T>(u)} { }

    template<Num U>
    constexpr void operator=(const U u) { val = convert_to<T>(u); }

    operator T() { return val; }    // extract value
};
```

Now we can simplify initializations and assignments by using **Number<T>** rather than **T** directly:

```
void test(int i)
```

```
        {
                Number<unsigned int> ii = 0;
                Number<char> cc = '0';

                ii = 2;         // OK
                ii = -2;        // throws
                cc = i;         // OK if i is within cc's range
                cc = -17;       // OK if char is signed; otherwise throws
                cc = 1234;      // throws if a char is 8 bits
        }
```

This is all that's required: 43 lines of code, most of which simply augment the type system and are never executed at run time. Those lines of code may be unfamiliar when first encountered, but they are not complex. What is done here is typically not done, implemented as costly run-time code, or provided as complex parts of a compiler.

An arithmetic type without arithmetic operations is of little use so we can add those:

```
        template<Num N1, Num N2>
        using Common = std::common_type_t<N1,N2>;   // a type that can represent a N1 and a N2

        template<Num I1, Num I2>
        auto operator+(Number<I1> x, Number<I2> y)
        {
                return Number<Common<N1,N2>>{ x.val + y.val };
        }

        template<Num N1, Num N2>
        auto operator*(Number<N1> x, Number<N2> y)
        {
                return Number<Common<N1,N2>>{ x.val * y.val };
        }

        // ... other operations, such as *, /, ==, and > ...
```

These operations determine the return type according to the rules of C++ using the standard-library concept **common_with**. Two types have a **common_with** type if there is type that they both can be converted into without loss of information. As usual, such predicates are evaluated at compile time and can be extended by users to handle user-defined types, such as **Bigint**.

As usual, signed/unsigned conversions are tricky. For example, **-1<2u** is false! The reason is that when an operation has a signed and an unsigned operand, the signed operand is converted to unsigned. Consequently, we define **<** (less than) for signed and unsigned integers to eliminate such horrors:

```
        template<Num N1, Num N2>
        bool operator<(Number<N1> x, Number<N2> y)
        {
                if constexpr (is_signed_v<N1> && is_unsigned_v<N2>)
```

```
            if (x.val < 0)      // x.val would be converted into a positive number
                return true;
        return x.val < y.val ;
    }
```

When using **Number**s, rather than built-in types directly, we outlaw all dangerous narrowing conversions. The run-time cost is very low because we only test where narrowing is possible. For example:

```
    Number<int> test(double d, Number<int> i)
    {
        auto x = d+i*10;                    // x is a double
        Number<double> z = d+i*10;          // OK
        return x*2 - d*z;                   // will check: may narrow
    }
```

# 3. Subscripting

Out of range accesses, often referred to as "buffer overflows" is a significant source of errors and security problems in C and old-style C++. An effective way of handling this problem is never to subscript a pointer without ensuring that the subscript is in range. That's impossible to do for arbitrary uses of pointers so we commonly use classes that check (and enforce the use of such classes). For example, a **span** is a type that holds a pointer to elements plus the number of elements it points to. The **span** doesn't own those elements; it simply controls access to them. A **span** has the information needed to prevent range errors and to implement range-**for** loops and algorithms on ranges (§4.3). It is a good example of a generic type. However, a **span** is still vulnerable to signed/unsigned problem when initialized with an explicit size. For example:

```
    const unsigned max = 100;
    int a[max];
    span<int> s {a, max-500};      // I mistyped 50
```

If a **span**'s constructor takes an **unsigned** size argument (like the standard-library **span**, but not the original GSL **span** [GSL]) that **-500** is converted to **unsigned**, so **s** gets a limit of **4294966896**. With that limit, no buffer overrun would be caught. With more complex size calculations, errors are easier to make and harder to spot in large programs.

## 3.1.  Protecting a span from narrowing

Let's address that problem using our **Number** and a variant of **span**. First decide what we want to be able to span over

```
    template<typename S>
    concept Spanable = ranges::contiguous_range<S>;
```

Contiguous ranges include C-style arrays, **std::array**s, and **vector**s. We name a **concept** even though we define it using a standard-library one because we might later want to extend it beyond the standard.

As a first step, let's define a span that can be initialized from data for which we know the number of elements:

```cpp
template<class T>
class Span {
    T* p;
    unsigned n;
public:
    unsigned check(unsigned nn)                        // in range?
    {
        if (n <= nn)
            throw Span_range_error{};
        return nn;
    }

    Span(Spanable auto& s) : p{ data(s)}, n{size(s)} {}   // [0:size)

    T& operator[](Number<unsigned> i) { return p[check(i)]; }
};
```

**Spanable<S>** (aka **ranges::contiguous_range<S>**) guarantees that **data(S)** and **size(S)** exist. This **Span** is easy to use:

```cpp
void test(Span<Number<int>> ssi, vector<double>& v)
{
    int x0 = ssi[10];          // OK if 10 is in range
    int x1 = ssi[-10];         // will throw

    Span<double> sv {v};
    int xx0 = sv[10];          // OK if 10 is in range
    int xx1 = sv[-1];          // will throw

    int a[100];
    Span<int> sa { aa };
    int xxx0 = sa[10];         // OK
    int xxx1 = sa[200];        // will throw
    int xxx2 = sa[sv[2]];      // will throw unless sv[2] is a positive integer
}
```

For **v**, we could have used a range-checked **vector** or checked the range ourselves using **v.size()**, but unfortunately people often fail to do that. C-style arrays turn into pointers at the slightest excuse and once that's done the number of elements pointed to is lost. A span carries the size along for later use.

## 3.2. Element type deduction

This **Span** is a bit verbose: we have to repeat the type of elements that the compiler already knows. We can avoid that by using what is called Class Template Argument Deduction (CTAD). We add a deduction guide to tell which part of an initializer's type to use as the template argument type:

```
template<Spanable R>
Span(R&) -> Span<ranges::range_value_t<R>>;          // use the range's element type
```

A lot of generic programming has to do with using information that the compiler already has. For arrays, the compiler knows its size and for **vector**s we know where to find it. That's what the **Span** constructor uses.

A Span **knows** its number of elements, so we can use range-for and other algorithms:

```
void test(Span<float> s)
{
    for (const auto x : s)
        cout << x << '\n';
}
```

That is far simpler and far safer than the common pointer and integer style

```
void test(float* p, int n)
{
    for (int i = 0; i< n; ++i)
        cout << p[n] << '\n';
}
```

However, a span must be more flexible to match what is necessary and frequently used: taking a sub-range. We can add constructors for that:

```
template<class T>
class Span {
    // …

    // initialize with a pointer and an explicit number of elements:
    Span(T* pp, Number<unsigned> nn) :p{ pp }, n{ nn } { }         // can't check

    Span(Spanable auto& s, Number<unsigned> nn)
        : p{ data(s) }, n{ size(s) } { n = check(nn); }            // [0:nn]

    Span(Spanable auto& s, Number<unsigned> low, Number<unsigned> high)
        : p{ data(s) }, n{ size(s) } { p=check(low); n=check(high - low); }  // [low:high]
};
```

We need appropriate deduction guides for the new constructors:

```
template<class T>
Span(T*, Number<unsigned>) -> Span<T>;

template<Spanable R>
Span(R&, Number<unsigned>) -> Span<ranges::range_value_t<R>>;

template<Spanable R>
```

**Span(R&, Number<unsigned>, Number<unsigned>) -> Span<ranges::range_value_t<R>>;**

Now we can write:

```
void test(vector<double>& v, int* p)
{
    int aa[100];
    Span s1 {aa};            // deduce Span to mean Span<int>
    Span s2 {aa,50};         // in case we want just half of the array
    Span s3 { aa,200};       // out of range: will throw

    Span sv1{v,10};          // first 10 elements
    Span sv2 {v,10,20};      // elements [10:20)

    int a[100];
    Span sa { aa, 10};       // first 10 elements
    Span sa { aa, 10, 20};   // elements [10:20)

    Span sp {p, 10};         // unchecked
}
```

If the container doesn't have the elements requested the constructor throws an exception. Now an explicit specification of a **Span**'s number of elements can often be tested and where it cannot (e.g., for a plain pointer) it stands out as an obvious target for code review and static analysis.

Given that we can deduce the type of element and sometimes even the number of elements for a **span**, we must wonder why we need to state the underlying type of a **Number**. We don't have to if we add type deduction:

```
template<Num Init>
Number(const Init) -> Number<Init>;    // deduce to the underlying type
```

Now **Number** is as notationally convenient as built-in arithmetic typers, but without their quirks:

```
Number x1 = 1;        // Number<int>
Number x2 = 1u;       // Number<unsigned>
Number x3 = 1.2;      // Number<double>
```

We need to be explicit only when that makes code clearer:

**Number<double> d = 1;**

## 4. Algorithms

Is what has been presented so far really Generic Programming? Much of what is commonly used to illustrate generic programming is mathematical code and data manipulation. The most popular C++ example is the standard-library framework of algorithms and containers (the STL). However, templates are the backbone of the ISO C++ standard library: time, random numbers, I/O, numerics, formatting, data manipulation, concurrency support, memory management, and more. That is, Generic Programming – programming with parameterized types – is excellent for managing the complexity of foundational high-performance code.

## 4.1. Classical sort example

To explore the use of concepts in generic algorithms, consider a popular example from the standard library: **sort**. Let's start with an old-style declaration that has been used since the 1990s and can still be found in the standard today:

```
template<typename Random_access_iterator>
void sort(Random_access_iterator first, Random_access_iterator last)
{
        // … implementation …
}
```

It says that **sort** takes **first** and **last** of some type **Random_access_iterator**. The standard states precisely what properties are assumed for **first**, **last**, and **Random_access_iterator** [C++20], roughly:

- **Random_access_iterator** must be a pointer-like type that can be used to iterate over a range of elements
- **Random_access_iterator** must provide random access to those elements.
- These elements must be comparable with the **<** operator.
- **first** and **last** must define a half-open range [**first**:**last**).

This style of definition offers flexibility and efficiency and was consequently spectacularly successful for decades. For example:

```
void test(vector<double>& vec, span<string>& ss)
{
        sort(vec.begin(),vec.end());      // vector of doubles in ascending order
        sort(ss.begin(),ss.end());        // span of strings in ascending order
}
```

However, this style of declaration does not meet the third design criteria for C++ generic programming (§1):

*Interfaces should be precisely specified in code so that humans and compilers can read, understand, and use them.*

Neither compilers nor most programmers read the standard and the declaration of **sort()** states only that some type is needed for its arguments. The result is that code that completely meets the expectations of the implementer works great but code that does not suffer spectacularly obscure compile-time error messages. For example:

```
void compute(vector<complex<double>>& v)
{
        sort(v.begin(),v.end());          // error: complex does not provide <
}
```

For most generic libraries, the specification of generic functions is less precise than the standard's and the error messages nowhere near as helpful as the comment in the **compute()** example.

## 4.2. Sort using concepts

Using concepts to specify the requirements of a template, we can do much better:

```
template<random_access_iterator Iter, typename Pred = ranges::less>
    requires sortable<Iter,Pred>
void sort(Iter first, Iter last, Pred p = {});
```

Here, the requirements on the argument types are explicitly specified:

- **Iter** must be an iterator providing random access to its elements (a **random_access_iterator**).
- **sortable<Iter,Pred>** checks that **Pred** can be used to compare of elements pointed to by **Iter**.
- **Pred is** defaulted to a standard-library less-than operation (**ranges::less**).
- If the caller doesn't specify a comparison, the default **Pred** (**Pred{}**) is used.

Given that, we can write:

```
void test(vector<double>& vec, span<string>&ss, list<int>& lst)
{
    sort(vec.begin(),vec.end());          // vector of doubles in ascending order
    sort(ss.begin(),ss.end());            // span of strings in ascending order

    sort(vec.begin(),vec.end(), ranges::greater{});   // descending order
    sort(ss.begin(),ss.end(), ranges::greater{});     // descending order

    sort(lst.begin(),lst.end());          // error: list doesn't provide random access
}
```

Types such as **vector**, **span**, and **list**, as well as algorithms like **sort**, **less**, and **greater**, and **concept**s such as **random_access_iterator** and **sortable** are defined in the standard library.

Compatibility is an important feature, so old code still works and with the same efficiency as before. However, we can now type check a call at the point of call rather than have to postpone checking until code-generation time (where error messages often are spectacularly bad).

## 4.3. Sort using ranges

The **sort** specified using concepts still doesn't check that a (**first**,**last**) pair really defines a sequence. For example, these kinds of bad errors have been observed in real-world code:

```
sort(vec.end(),vec.begin());        // begin and end flipped
sort(ss1.begin(),ss2.end());        // not iterators into the same container
```

Furthermore, in 90+% of cases, specifying **begin** and **end** is verbose and not really what we wanted to say. We want to sort a whole container. The obvious solution is to give **sort** a range, rather than a pair of iterators supposedly representing a range. First, we define what we mean by a sortable range:

```
template<typename R, typename Pred = ranges::less>
```

```
concept Sortable_range =
        ranges::random_access_range<R>              // has begin()/end(), size(), ++, [], +, …
        && sortable<ranges::iterator_t<R>, Pred>;   // compare elements using Pred
```

That is, to be a **Sortable_range**, a type must

- be a range providing random access
- provide an iterator that provides operations necessary for sorting (such as **swap**) and allows elements to be compared (using a predicate **Pred**).

Given **Sortable_range**, we can define a further improved **sort**:

```
template<typename R, typename Pred = ranges::less>
        requires Sortable_range<R,Pred>
void sort(R& r, Pred p = {})
{
        sort(r.begin(), r.end(), p);
}
```

Now we have what I consider a minimal and elegant interface to whatever **sort** we want:

```
void test(vector<double>& vec, span<string>&ss, list<int>& lst)
{
        sort(vec);      // vector of doubles in ascending order
        sort(ss);       // span of strings in ascending order
        sort(lst);      // error (as ever): list doesn't provide random access

        sort(vec, ranges::greater{});   // descending order
        sort(ss, ranges::greater{});    // descending order
}
```

## 4.4. Function template overloading

Given the information in concepts, function overloading is trivial. If we wanted to, we could define **sort** for **list**s. First, we define the requirements for a **sort** that doesn't require random access:

```
template<typename R, typename Pred = ranges::less>
concept Forward_sortable_range =
        ranges::forward_range<R>                    // has begin()/end(), ++, …
        && sortable<ranges::iterator_t<R>, Pred>;   // compare elements using Pred
```

There are clever algorithms that sort streams of elements without requiring random access (e.g., **std::list::sort()**), but for many uses we could just copy the elements of the **list** into a **vector**, sort the **vector**, and copy the elements back into the **list**.

```
template<typename For, typename Pred = ranges::less>
        requires Forward_sortable_range<For, Pred>
void sort(For & r, Pred p = {})                     // one possible implementation
{
```

```
            vector v {from_range, r};      // copy the elements into a vector
            sort(v, p);                    // use the vector sort
            copy(r, v.begin());            // copy the element back
    }
```

Now we can write:

```
    void test()
    {
            vector<double> vec = {1, -2, 2, 3};
            list<string> lst = {"d", "q", "a"};

            sort(vec);                     // sort the vector ascending
            sort(lst,ranges::greater);     // sort the list descending
    }
```

Whether sorting **list**s without explicit mentioning "list" is wise is debatable. In particular, the reason the C++ standard doesn't support sorting **list**s is that the memory layout of **list**s is inappropriate for fast sorting.

Fortunately, the overloading rules for template functions are very simple:

- if no function matches, the call fails (obviously).
- If just one function matches, it is used.
- If two functions both match based on their concepts and one has constraints that are a subset of the other's, the one with the stricter requirements is chosen.
- Otherwise, the call is ambiguous and an error.

One important use of overloading is to allow tuning by providing specialized data structures and algorithms for performance critical tasks. Concepts significantly simplify this by selecting the optimized versions at compile time.

## 5. Concepts

Concepts are simply compile-time functions (predicates) that can take type arguments. They are used to provide significant flexibility to the type system. Like other compile-time functions, they can be used to catch errors early and to move computations from run time to compile time. Importantly, that saves us from writing error-handling functions.

We have always had the notion of concepts. For example,

- **C built-in types**: arithmetic and floating (from 1972 or so [K&R'78])
- **STL concepts**: iterators, sequences, and containers (since the early 1990s [AS'95])
- **Mathematical concepts**: monad, group, ring, and field (for a couple of centuries)
- **Graph concepts:** edge, vertex, graph, and DAG (since 1736)

No generic program could work unless the programmer had an idea of the concepts involved clearly in mind. What is relatively new [BS'17, C++20] is that we can define them to be used in code.

Concepts are not types of types or defined properties of types (like base classes). They are functions used to inquire about properties of types. For example, "are you an iterator?", "are you a number?", and "can this range together with this comparison predicate be used by **sort**?" These functions are executed at compile time. That makes them very efficient in addition to the usual properties of functions: general, flexible, and easy to use. Importantly, concepts specify what a template must be able to do with its arguments, not exactly what an argument type must provide to do so.

Many, probably most, concepts take multiple arguments. Like many "ordinary functions", concepts tend to be defined by calls to other concepts. We saw that with **Forward_sortable_range**:

```
template<typename R, typename Pred = ranges::less>
concept Forward_sortable_range =
    ranges::forward_range<R>                    // has begin()/end(), ++, …
    && sortable<ranges::iterator_t<R>, Pred>;   // compare elements using Pred
```

It takes two type arguments (**R** and **Pred**) and calls two other concepts using those (**ranges::forward_range** and **sortable).** To be able to relate **R** and **Pred** to each other, **sortable** takes both as arguments. If the comments are not sufficiently informative for use, the standard and online sources (e.g., [CPP]) have detailed definitions.

## 5.1. Use patterns

When we want to express something that hasn't already been defined by others, we define concepts directly from language features. The classical example is how to require that a pair of types can be compared using **==** and **!=**:

```
template<typename T, typename U = T>
concept equality_comparable = requires(T a, U b) {
    {a==b} -> Boolean;
    {a!=b} -> Boolean;
    {b==a} -> Boolean;
    {b!=a} -> Boolean;
}
```

Here, we use a **requires**-clause taking two arguments. Its arguments are used only for type checking (and never executed at run time). The expressions in curly braces are *use patterns* [GDR'06]. That is, expressions must type check yielding a result of the concept mentioned after the **->** (if any). Therefore, **{a==b} -> Boolean** says that for **T** and **U** to be **equality_comparable**, the expression **a==b** must yield a type that is **Boolean** (e.g., a **bool**). Fortunately, the standard library provides an **equality_comparable** so we don't have to define it ourselves, but the definition above shows the basic technique for defining a **concep**t from fundamental language features.

A **requires**-clause building directly on language features is a low-level mechanism for checking whether a construct is valid C++. It is essential for expressing low-level requirements but best avoided outside **concept** definitions. Named concepts (often defined using **requires**) are more comprehensible and maintainable. Outside concept definitions, **requires**-clause are ad hoc, and programmers often fall into the trap of constraining the actual implementation of a function rather than presenting a general idea as a concept. This easily leads to many similar but slightly different **requires**-clauses, rather than a concept that can be used repeatably and is more easily remembered. In other words, overuse of **requires**-clauses degenerates into copy-and-paste programming.

A concept specifies what a template must be able to do with its arguments, not exactly what those argument types must be. For example: when specified in a use pattern, the **+** in **a+b** for some type **T** and some type **U** could be provided as any of

- **X operator+(X,Y);**                          // *if a is an X and b is a Y*
- **X X::operator+(const Y&);**                  // *if a is an X and b is of a class derived from Y*
- **Y operator+(const X&, const Y&);**           // *if an X can be implicitly constructed from a T*
                                                 // *and a Y can be implicitly constructed from a U*
- **Y operator+(Y,X&);**                         // *if a Y can be implicitly constructed from a T*
                                                 // *and b is an X*
- *… and many more …*

That's important because use patterns allow concepts to

- Handle mixed-mode arithmetic
- Handle implicit conversions
- Provide interface stability (e.g., if the definition of a **+** changes or if a new **+** is added)

## 5.2. Checking types against concepts

It is not necessary for users to explicitly check whether a type meets the requirements of a concept. That is implicitly done whenever such a combination is used. That implies that combinations of types and concepts that a programmer hasn't imagined can be safely used and provide an important degree of flexibility. However, if we want to catch problems early for combinations we know of (e.g., during debugging or testing), we can test that types are **equality_comparable** using **static_assert**:

```
static_assert(equality_comparable<int,double>);    // succeeds
static_assert(equality_comparable<int>);            // succeeds (U is defaulted to int)
static_assert(!equality_comparable<int,string>);    // succeeds (Note the !)
```

Concepts provides a cheap and very flexible mechanism that enables us to build our own type systems on top of what C++ offers by default. A single-argument **concept** can be used almost like type and multiple-argument **concept**s can define relationships among types.

# 6. Facilities supporting generic programming

Every C++ feature can be used in generic code. Some language features exist primarily to support generic programming, but these features are designed to combine nicely with other language features, so you find them used in ways that are not primarily generic programming. The idea is to avoid generic programming becoming a sub-language isolated from the rest of C++.

## 6.1. Uniform treatment of types

From the earliest days, C++ treated user-defined types (e.g., **vector** and **complex**) and built-in types (e.g., **int** and **double**) uniformly with respect to scope and definition syntax. That's necessary to avoid special cases in generic code. Consider a simplified version of **swap()**:

```
template<typename T>
void swap(T& x, T& y)
{
    T tmp = std::move(x);
    x = std::move(y);
    y = std::move(tmp);
}
```

If user-defined types needed to be created with **new** (as in Simula and other object-oriented languages) this would require a workaround for common types like **complex** and **vector**.

The use of constructors/destructor pairs (the oddly named "Resource Acquisition Is Initialization" (RAII) idiom) allows us to avoid special cases for objects that require cleanup or resource release. Consider a simplified version of **make_unique()** that constructs an object of an arbitrary type and returns a unique pointer to it:

```
template<typename T, typename... Args>        // variadic template (see §6.5)
auto make_unique(Args&&... args)              // the return type is deduced
{
    return std::unique_ptr<T>(new T(std::forward<Args>(args)...));
}
```

If built-in types couldn't be created using **new** and late destroyed using **delete** by **unique_ptr**'s destructor, this would require a workaround. If "cleanup operations" couldn't be made implicit defining destructors, **unique_ptr** would need workarounds for every type requiring cleanup.

## 6.2. Lambdas

We used the standard-library function object **greater** to reverse the order of the sorted **vector**: **sort(lst,ranges::greater)**. For a predicate that has many uses, represents a general action, and can be given a comprehensible name, that's ideal. However, many arguments to algorithms don't fit those criteria and having to define the function object (a class with an **operator()**) in one place and use it in another can be a burden to the programmer. Consequently, C++ offers a notation for specifying an unnamed function object that can be used immediately. For example:

> **sort(lst,[](const string& x, const string& y) { return x>y; })**.

The **[]** indicates an anonymous function object. Such objects are called "lambda expressions" or simply "lambdas." Here, the **[]** simply generates an anonymous function.

A lambda can have a state and access local variables in its enclosing scope. For example:

```
void test(vector<int>& v, int smallest)
{
    auto p = find_if(v, [&](auto arg) { return smallest <= arg; });
    if (p!=v.end()) {
        // … use *p …
    }
}
```

Here, the lambda holds a reference to the local variable **smallest** and uses it as the smallest value to be found. The writer of a lambda can exercise detailed control over what names from its environment it can access ("it captures") and whether their use is by reference or by copy. "By copy" is especially useful when we want to minimize indirections (through pointers and references) as in much concurrent programming. The **[&]** indicates "by reference."

In the lambda passed to **find_if** above, we used plain **auto** as the parameter type. That means that the lambda will accept any type of argument. We could get away with that here because we know that **find_if** only calls that argument with an argument of its element type and that **v**'s element type is **int**. But what if we didn't know that much and made a mistake? We would get a late and poor error message from the compiler. To avoid that possibility we could use a more restrictive concept:

```
auto p = find_if(v, [=](Num arg) { return smallest <= arg; });
```

While cleaning up that call, we changed the capture of **smallest** from by-reference to by-value. If that lambda is called many times, that's more efficient and when capturing by-value we eliminate the possibility of making the mistakes of retaining a reference when we should not.

## 6.3. Generating types

A generic algorithm wouldn't be of much use unless there was a set of types it could be generic over. A key use of templates is to generate types with similar interfaces. As usual, **vector** offers a simple example:

```
template<Element T, Allocator A = std::allocator<T>>
class Vector {
        A alloc;
        T* elem;
        int sz;
public:
        Vector(int sz, const T& val = {}, Alloc a = {})
                : alloc{a}, elem{alloc.allocate(sz)}, sz{z}
        {
                for (auto x : span{elem,sz})
                        alloc.construct(x,val);
        }

        ~Vector()
        {
                for (auto x : span{elem,sz})
                        alloc.destroy(x);
                alloc.deallocate(sz);
        }

        // … access operations …
};
```

This is a much-simplified version of the standard-library **vector.** We used an **Element** concept to represent the requirements for a type to be used for elements, notably the ability to be copied and moved. The concept **Allocator** represents the requirements for a type to be used to allocate and deallocate memory. The allocator allows the elements of a **Vector** to be placed in the free store (dynamic memory, heap), on the stack, in a user-define storage pool, or even in a permanent memory. Both the allocator (**alloc**) and the default element value (**val**) are defaulted to their type's default value. The standard-library default allocator uses **new** and **delete**.

## 6.4. Optional member functions

When defining a parameterized type, it is often useful to provide an operator for some but not all template arguments. For example, a "smart pointer" should offer a **->** operator if and only if it points to a class object with members:

```
template<typename T> class Ptr {
    // …
    T& operator*();
    T* operator->() requires is_class<T>;    // offer -> (only) if T is a class
};
```

Now, we get

```
void test(Ptr<int> pi, Ptr<pair<string,int>> psi)
{
    auto x0 = *pi;              // OK: x0 is an int
    auto x1 = *psi;             // OK: x1 is a pair<int,string>

    auto x2 = pi->value;        // error: Ptr<int> doesn't have a ->
    auto x3 = psi>first;        // OK: x3 is an int
}
```

Such optionality is quite popular. For example, the standard-library **pair** type offers a constructor from another **pair** provided each element can be converted:

```
template<typename T, typename U>
struct Pair {
    T t;
    U u;
public:
    // …
    // offer constructor (only) for types that can be converted to the members:
    template<typename TT, typename UU>
        requires convertible_to<TT, T> && convertible_to<UU, U>
    Pair(const Pair<TT, UU>& pp) :t(pp.t), u(pp.u) {};
};
```

The **convertible_to<TT,T>** is a standard library concept checking that a **TT** can be implicitly converted to a **T**. For example:

```
void test(Pair<int, int> pii, Pair<double, double> pdd, Pair<string, int> psi)
{
    Pair<int, int> x0 = pii;              // OK
    Pair<int, int> x1= pdd;               // OK
    Pair<double, double> x2 = pii;        // OK
    Pair<int, int> x3 = psi;              // error:  can't convert first element
    Pair<string, int> x4 = psi;           // OK
}
```

However, **convertible_to<TT,T>** doesn't protect against narrowing conversions (e.g., **x1**), so it might be better to simply check if the values used really results in narrowing using **convert_to()** from §2.1:

```
template<typename TT, typename UU>
Pair(const Pair<TT, UU>& pp)
        :Pair{convert_to<T>(pp.t), convert_to<U>(pp.u)} {}
```

This catches more errors but incurs a slight cost when initializing from a **Pair** type that potentially narrows.

We left a problem when we defined **convert_to** only for numeric types, so conversions involving **string**s now fail. That now seems restrictive. A simple solution to this is to overload **convert_to** for types that are not numeric so that we can't check for narrowing:

```
template<typename T, typename U>
constexpr T convert_to(U u)
{
    return T{u};
}
```

Thanks to overloading (§4.4), this less strict version will be chosen for types that don't meet our definition of numeric types from §2.1, such as strings, pointers and enumerations. This version uses the **T{u}** notation to protect from pointer-to-**bool** and **int**-to-enumeration narrowing conversions. Such narrowing conversions now trigger compile-time errors.

## 6.5. Variadic templates

Sometimes, a function needs to take a variety of arguments. For many simple cases, this can be achieved through overloading (e.g., §4.4) or default arguments (e.g., §4.3) but C++ also offers a general mechanism that can handle a varying number of arguments of a variety of types.  For example:

```
print1("Hello ", "world", '!', " It's now ", chrono::system_clock::now());
```

Just now that produced

> **Hello world! It's now 2025-05-22 16:55:25.2750128**

This can be achieved by something called variadic templates:

```
template<class T>
concept Printable = requires(T t, ostream & os) { os << t; };
```

```
template<Printable T>
void print1(T... args) { (cout << ... << args); }
```

First, we define a concept **Printable** as simply something that can be written to an **ostream** using **<<**.

Next, we define a function that takes a sequence of zero or more **Printable** arguments. The ellipsis (…) indicates zero or more values and such an argument is called a parameter pack.

The simplest way of using a parameter pack is in a fold expression. Here, **(cout << ... << args)** says "write all the elements of **args** to **cout** one at a time using **<<**." Basically, we say **<<…<<** instead of **<<** to indicate that a parameter pack is used instead of a single object.  A fold expression can be done with most operators.

If we want to do more with a parameter pack than can be handled by a single operator, we have to recursively unravel the pack. For example, we can implement something similar to the standard library **print** function [C++20, BS'22]) that uses a format string where **{}** indicates where arguments should be inserted. For example:

```
print2("Hello {}! It's now {}", "world", chrono::system_clock::now());
```

That produced:

**Hello world! It's now 2025-05-22 17:50:42.3606077**

Implementing **print2** involves classical C-style character manipulation in addition to variadic templates and **concept**s.

First, we handle the case with no arguments after the format string:

```
void print2(const char* s)
{
        if (s == nullptr)
                return;
        while (*s) {
                if (*s == '{')
                        if (*++s == '}')
                                throw runtime_error("argument missing");
                        else
                                cout << '{';
                cout << *s++;
        }
}
```

Most of this code is there to catch the error when the format string requires an argument that isn't supplied.

Next come the generic programming solution using a parameter pack (**T…**). First, we deal with the first element of the pack and then recursively call **print2** again with whatever is left of the format string and the rest of the arguments:

```
template<Printable T>
void print2(const char* s, T val, T... args)
{
    while (s && *s) {
        if (*s == '{')
            if (*++s == '}') {
                cout << val;
                return print2(++s, args...);         // handle the rest
            }
            else
                cout << '{';
        cout << *s++;
    }
    throw runtime_error("too many arguments");
}
```

When **args…** is empty, the one-argument **print2** is called. That way, we never get to the **"too many arguments"** error unless there really too many arguments.

## 6.6. Static reflection (C++26)

A lot of the power of generic programming comes from letting the programmer use what the compiler knows about the types and functions. Static reflection is a facility coming in C++26 that will dramatically enhance the programmer's ability to use what the compiler knows.

As an example, let's generate a description of a class' layout. Various forms of such layout data can be used in a variety of places, such as debuggers, cross-language interfaces, and communication protocols:

```
struct member_descriptor
{
    string_view name;
    size_t offset;
    size_t size;
};
```

We use a **string_view** (a type similar to a **span** for character strings) to avoid creating a **string** object. The compiler already has such a string stored for the lifetime of the type we are examining.

We want to make a **member_descriptor** for a variety of types. Here is a function doing that:

```
template <typename S>
consteval auto get_layout()
{
    constexpr auto members = meta::nonstatic_data_members_of(^^S);

    array<member_descriptor, members.size()> layout;

    int i = 0;
    for (const auto& x :members)
        layout[i++] = {
```

```
                    meta::identifier_of(x),       // member name
                    meta::offset_of(x),
                    meta::size_of(x)
            };

        return layout;         // returns an array<member_descriptor, members.size()>
    }
```

The **^^S** means "get the reflection object of type **meta::info** for **S**." A **meta::info** object gives access to essentially all the compiler knows about a type and offers compile-time functions to access it. Here, **nonstatic_data_members_of(^^S)** returns a (fixed sized) **std::array** of descriptors of **S**'s members. We can then access the properties of those members, here their identifiers, offsets, and sizes.

As demonstration, we can do a small test. Source code:

```
struct X {
    char a;
    int b;
    string c;
};

constexpr auto Xd = get_layout<X>();
```

Now **Xd** is an array of **member_descriptor**s with the value **{{"a", 0, 1}, {"b", 4, 4}, {"c", 8,2 4}}**. The literal strings are what you'd see looking through the **string_view**s. Naturally, the exact values of the offset and size depend on the implementation.

This style of code can be used for generating I/O operations, serialization to/from various formats, foreign language calls to/from C++, and much more. It is too early to firmly predict how static reflection will be used in general, but it obviously has a major role to play for experienced programmers implementing functions and data structures with conventional interfaces that can be easily used. C++26 doesn't allow us to directly inject functions into code, but if we need to, we can always output the source code of the function as a string and then compile such strings in a different translation unit.

# 7. Major concepts design decisions

The development of generic programing support since the earliest days of C++ is well documented [BS'82, BS'94, BS'03, BS'03b, AS'09, BS'09, AS'11, BS'20]. This section briefly presents key design decisions related to concepts; that is, to the use of requirements that code can place on the types and functions it uses.

My ideal is for generic programming to be simply programming without any special restrictions or separate rules. This is of course not exceptional (e.g., see ML [LP'91]) but given generic programming's gradual and slow introduction into C++, that ideal is not widely understood or generally appreciated. However, as seen above, we can often approximate this ideal. This section presents some rationale for key design decisions for **concept**s.

## 7.1. Concepts are functions

Many constraints systems rely on sets of functions, much like a class definition. The design of concepts for C++0x was an example of that [DG'06]. Such systems suffer from rigidity as the type of each constraint function (argument and return types) must be specified. Also, the constraints of each function argument must be specified in isolation from others. This leads to problems specifying implicit conversions and to problems specifying constraints on operations with multiple operands of different types.

Using compile-time-evaluated functions and use patterns (§5) solve those problems, make almost the complete language available for specifying requirements, simplify such specifications, and provide a direct mapping from requirements to compile-time checking. Concepts specify what a template must be able to do with its arguments, not how the argument types must be defined to do so. In particular, concepts handle mixed-mode arithmetic and implicit conversions simply and elegantly (§2.2, §5.1).

A 2006 paper [GDR'06] introduced the notion of specifications as use patterns but unfortunately the community found it too unusual and set concepts back about 15 years. See §5, §7.1, and §7.4 for some reasons for preferring use patterns over the conventional class-like specifications of requirements.

## 7.2. GP and OOP

For C++, generic programing (GP) was designed to complement classical Object-Oriented Programming (OOP). Both are mechanisms for specifying abstractions to represent entities and operations in an application. GP and OOP each have unique strengths not shared with the other. Thus, despite overlaps in usage, they are complementary rather than complete alternatives.

GP offers what is often called static polymorphism, relying on compile-time selection of functions. This is in contrast to the run-time polymorphism offered by OOP, relying on class hierarchies using virtual functions to select the appropriate derived class.

GP is focused on functions (algorithms). In particular, it focuses on a function's requirements on its arguments, including the relationships among those arguments; that is, on concepts.  An argument list can involve any set of types and values (§7.5) that meet the function's concepts. A non-generic function is simply one that requires a specific set of types.

In contrast, OOP is focused on objects. In particular, it focuses on the interface to and representation of individual objects; that is, on classes. The implementation of the interface can be refined through derived classes overriding virtual functions while keeping the meaning of operations unchanged (interface inheritance).  Functions relying on multiple arguments of classes with virtual functions (e.g., an intersect operation of shapes) can be tricky (double dispatch [BS'94], multimethods [PP'09]). If an object of a derived class cannot be used in place of one of its base classes, the inheritance should be **private** (this has been in C++ from the start [BS'82] and was later named "substitutability"). A class need not be part of a class hierarchy and need not have virtual functions.

GP enables inlining of operations on function arguments because the definition of the types of those arguments must be in scope. This offers great opportunities for optimization that compilers take advantage of, including optimizations involving inlining of operations involving combinations of argument.

In contrast, OOP often relies on virtual function calls where the actual type of an object is not known until run time. Only calls to non-virtual functions, functions declared **final**, and functions of objects whose definitions are in scope can be inlined.

In OOP, we have to pre-define interfaces as base classes of classes organized into hierarchies. That required foresight – sometimes a great deal of foresight – from the designer of a class hierarchy to avoid workarounds or widespread code changes if/when a base class needs a change.

In contrast, concept-based GP enables the use of a set of argument types as long as it meets the code's requirements (§4.4). That's more flexible and requires less foresight from the programmer than class-hierarchy-based OOP but doesn't cope with objects of types that are not known until run time.

In OOP, an object accessed through a class interface can be of various different derived classes of different sizes. Consequently, objects are typically allocated on the free store (dynamic memory, heap) and accessed indirectly through pointers or references. This then requires that more care is to be taken with memory management (e.g., using **unique_ptr** or **shared_ptr** to avoid resource leaks and dangling pointers). Also, indirection can impede optimization and can mess up cache usage.

GP and OOP can – and often are – used in combination. GP can use OOP constructs and techniques in their implementation and use pointers and references to base classes as arguments. Similarly, OOP can use GP techniques in their implementation. However, parameterizing a class that has virtual functions should be done only with great care because that can lead to code bloat from virtual functions that are not actually used.

Let's have a look at a generic version of the classical "draw all shapes" OOP example:

```
void draw_all(Drawable_range auto& r)
{
        for (auto& d : r)
                r->draw();
}
```

It is clearly OOP. Its simplest use involves giving it a **vector<Shape*>** as an argument and relies on virtual function calls. It is clearly also a generic program, because **draw_all** is a template that we could also pass a **list<unique_ptr<Shape>>**. It can even handle a type that controls access to **Drawable** objects of a variety of types that aren't in a hierarchy as long as they all have a **draw()** function that can be accessed through a **->** operator.

A very simple version of **Drawable_range** looks like this:

```
template<typename T>
concept Drawable = requires(T a) { a->draw(); };

template<typename R>
concept Drawable_range = ranges::forward_range<R> && Drawable<ranges::iterator_t<R>>;
```

## 7.3. Notation

C++ is an old language, parts stretching back all the way to the early C (1972 [K&R]). The support for generic programming gradually evolved from the earliest days [BS'82] over decades under the constraints of the language at different stages of development and the varying tastes and opinions of a large standards committee.

I chose the **< … >** notation for type parameterization following some use in theory. Initially, I did not use the prefix **template** keyword: **< … >** was a suffix to the name they parameterized. However, people strongly insisted on having a prefix keyword to make templates stand out. That is typical, initially people ask for a LOUD syntax for novel constructs because they are seen as difficult or even dangerous. Later, the same people complain about verbosity.

The same happened when I proposed concepts to be usable with exactly the same syntax as types. For example:

    **Void sort(Sortable_range& r)**         **//** *conventional functional syntax*

as an alternative to

    **template<Sortable_range R>**         **//** *shorthand syntax (C++20)*
    **void sort(R& r);**

or even

    **template<typename R>**         **//** *explicit qualification syntax (C++20)*
        **requires Sortable_range<R>**
    **void sort(R& r);**

In the end

    Void **sort(Sortable_range auto& r);**     **//** *almost functional syntax (C++20)*

was accepted as a compromise. The last three notations are supported by the standard. There are cases where the longer notations are needed, especially to handle multi-argument functions and functions where the name of the argument type is needed in the definition code. Where they can express the same constraint, the notations are interchangeable. Basically, **<typename T>** is the mathematical "for all types **T**" and **<Foo T>** for a **concept Foo** is the mathematical "for all **T** such that **Foo<T>**."

The name "concept" was coined by Alex Stepanov in 1981 because concepts are intended to represent fundamental concepts in code [DK'81, AS'09]. It thus predates most names for constraints.

The name **auto** is part of the C++ syntax and used to represent any type. I suggested it in 2002 as the start of a push towards more support for generic programming, e.g., allowing **auto f(auto)**. It was accepted in C++11 [BS'02, JJ'03]. I see **auto** as the weakest, least restricted concept. We could now do without it and use a defined **concept** instead:

    **template<typename T> concept Auto = true;**

Now **Auto** accepts values of every type just like **auto**.

## 7.4. Concepts can be partial constraints

Consider a concept specified directly from language primitives:

```
template<typename T, typename U = T>
concept Arith = requires(T x, U y) {      // arithmetic operations and a zero
        x+y;    x-y;    x*y;    x/y;
        x+=y;   x-=y;   x*=y;   x/=y;
        x=x;                              // copy of a T  (not x=y)
        x=0;
};

template<typename T, typename U = T>                    // symmetric
concept Arithmetic = Arith<T,U> && Arith<U,T>;
```

This **Arith** concept accepts any pair of types **T,U** with the operations specified but will often be used with a single type. The **Arithmetic concept** additionally requires those operations for the types reversed: **U,T**. For example:

```
template<typename T, typename U>
double my_computation(Arith<T> auto x, Arith<U> auto y)
        requires Arithmetic<T,U>
{
        // …
}
```

Return type constraints are a bit tricky, so we left them out. Incomplete concepts are very useful because they catch most errors early. They are inevitable during development where we have yet to discover all uses. Importantly, errors not caught early are, as ever, caught at template instantiation (code generation) time. Then, we get awful error messages but also the opportunity to improve the relevant concept.

## 7.5. Concepts can take value arguments

Some constraints on generic code involve values in addition to types just as other templates. For example, we can specify that a generated type needs a minimum size that is a power of two:

```
consteval bool is_power_of_two(int n)
{
        return 0<n && (n & (n - 1))==0;
}

template<int S>
concept Buffer_space = (1024 <= S) && is_power_of_two (S);

template<Element T, int S>
        requires Buffer_space<S>
struct Buffer {
        T buf[S];
        // …
```

```
    };

    void test0()
    {
            Buffer<char, 100> b1;        // error: buffer too small
            Buffer<int, 10000> b2;       // error: size not binary
            Buffer<int, 2048> b3;        // OK
    }
```

Such value arguments can be most useful. One reason is that violations are detected at compile-time so that we don't have to write an error handler.

## 7.6. Concept type matching

Whenever we look up a name, there is a chance of finding something that didn't match our expectations, a false match. That's the case for libraries, for nested scopes, and for class hierarchies. When looking for typed entities (especially functions), the chance of a false match is limited by type checking. When looking for a name constrained by concepts, the chance of a false match is limited by the concept check.

Before concepts, that check was missing for template name lookups and problems occurred because only the template name was involved in the search and constraints were only applied (too) late, at instantiation time. The result was obscure error messages and occasionally "the wrong" template was found (just as occasionally an unexpected function is used after a library lookup). Concepts dramatically change that. Most well-designed concepts involve a whole set of requirements, rather than a simple type check. For example, consider **Arithmetic** from §7.4, which requires 10 operations. A type that doesn't represent a number is most unlikely to match all those. Even a simple concept like **Printable** from §6.4 requires as much checking as a traditional type check. Thus, concept checking gives about as much protection or more against false matches as traditional type checking.

There are concepts that differ only semantically and thus cannot be distinguished syntactically without special effort. For example, the standard-library **forward_iterator** concept differs semantically from an **input_iterator** only in that a **forward_iterator** allows repeated traversals of its sequence. The simple solution to such problems is to introduce a syntactic difference indicating the semantic requirement. For example, a **forward_iterator** must have a **forward_iterator_tag** and an **input_iterator** must not.

## 7.7. Definition checking

Like most people, I initially thought that it was important for generic code to be checked in isolation relying exclusively on the concepts specified [D&E, BS'02b]. I now consider that a mistake because isolating templates decreases their usability and can have negative performance implications. So, C++ concepts don't offer definition checking even though we know how to implement it if we wanted to [GDR'12].

Consider:

```
    void advance(input_iterator auto p, int n)
    {
            log("advance({},{})(",p,n);       // Should this work?
```

```
        p+=n;                        // Should this work? Where to check?
    }
```

Here, the interface to **advance** doesn't mention the logging function **log**. Should a name unmentioned in the interface be usable from a template? If yes, we need to change the interface to be able to use **log**. That would imply interface instability and code changes. If not, we cannot compile templates in isolation.

In industrial code, support for infrastructure, monitoring, telemetry, logging, debug aids, and the like are very common. Such support code changes over time and have nothing to do with a function's primary aim so they should not be part of its interface.

Also, **p+=n** wouldn't compile because an **input_iterator** doesn't support **+=**. However, many **input_iteraror**s are also **random_access_iterator**s, so banning **p+=n** would also have performance implications. Consider:

```
vector v = {0,1,2,3,4,5,6,7,8,9};
auto p = v.begin();
advance(p,2);                 // Should this work? It has since C++98
```

The fact that **advance(p,2)** works even though **input_iterator** doesn't offer **+=** is important for performance. If it didn't, the performance of some standard-library algorithms implemented using **advance** would change from O(n) to O($n^2$).

What happens is that a final check of **advance**'s definition is postponed until the actual type of the iterator is known. That is necessary anyway if we want to avoid potentially costly indirect function calls in the implementation.

## 8. Language design challenges

C++ is deliberately an evolving language [BS'94]. It is thus entirely expected that its support for generic programming is incomplete and improving. We learn through feedback from use. Consider a few areas that need exploring further in the context of concept-based generic programming because they appear to promise significant improvements:

- *Axioms*: Like most languages, C++ does not have a way to specify semantic properties of constructs for the use of analyzers and code generators. A design was accepted for C++11 [GDR'09] but was removed again with C++11 concepts [BS'09] and not reintroduced for lack of committee time.
- *Output ranges*: To increase range safety, it is necessary to be able to range check output operations. This is easily done [BS'22], but for systematic and wide use, standard-library support is needed.
- *Notation*: The use of **auto** after concepts (e.g., **Drawable_range auto** (§7.2)) is redundant in almost all cases and distracting. To make the conventional functional notation with concepts more expressive, we might introduce a way of naming the deduced type [MS'23].

- **Overloading of classes**: Allowing overloading of classes based on concepts, as is done for functions, would provide a generalization of and possibly a simplification of what is currently done with specialization.
- **Pattern matching**: Functional-programming style pattern matching would enable selection based on concepts as well as types and values. That would simplify much code and eliminate some opportunities for type violations. As for functions, the notation for pattern matching must be designed to minimize the distinction between types and concepts [HS'24].
- **Uniform function call**: the difference between conventional functional style notation for function call (e.g., **f(x,y)**) and OOP style (e.g., **x.f(y)**) is a distraction and leads to duplication in concepts. It is also unnecessary [BS'15].

None of this is technically difficult but may be politically impossible in a large standards committee.

## 9. Summary

Generic Programming is "just" programming, grounded in classical mathematics [AS'95, AS'14]:

*"The most general, most efficient, most flexible representation of concepts"*

Contemporary C++ approximates this ideal quite well. As examples, we show how to eliminate signed/unsigned conversion problems (§2) and common range errors (§3). Generic programming is the backbone of the ISO C++ standard library (§4).

Generic programming is complementary to classical Object-Oriented Programming (§7.2).

Constraints on generic algorithms are expressed as concepts. Concepts

- Are compile-time functions that can take multiple type and value arguments – a concept is not specified as a class with sets of function declarations nor (just) a type of types (§5, §7.1).
- Can be defined in terms of other concepts or directly in terms of basic language constructs using use patterns (§5).
- Specify what a generic function requires from its arguments – not how those arguments implement operations on arguments (§7.1).
- Can be used to select among alternatives through simple and flexible overloading (§4.4, §7.1).
- Allow us to make generic programming very similar to "ordinary" (non-generic) programming without adding run-time overheads (§2, §4.4, §7.3).

Generic programming using concepts is deeply integrated into C++, rather than being an isolated sub-language. For example:

- Built-in and user-defined types are treated uniformly, as needed to avoid generic code having to distinguish between those (§6.1).
- Implicit cleanup and resource release (RAII) saves us from dealing separately with types that require cleanup or resource release from types that do not, and from having to litter code with error checks for exceptional errors (§6.1).
- Lambdas allow operations as arguments to generic code to be specified exactly where they are needed (§6.2) and lambdas can carry state.

- Parameter packs allow us to pass lists of arguments of differing types (§6.5).
- Static reflection (C++26) greatly increases the ability of implementers of generic code to use information known to the compiler (§6.6).

Generic programming using concepts and other compile-time evaluation support enable us to build our own type systems on top of what C++ offers by default (§2.2, §5.2).

Stroustrup 2025                                          Concept-Based Generic Programming

# 11. Acknowledgements

I am grateful to Chuck Allison, Alfred Bratterud, Damask Talary-Brown, James Cusick, J. Daniel Garcia, Peter Juhl, Ole Lehrman Madsen, Arne Tolstrup Madsen, Gabriel Dos Reis, Herb Sutter, Clifford Tiltman, Ville Voutilainen, and J.C. van Winkel who provided constructive comments on earlier versions of this paper – especially helping to make the ideas comprehensible to a wider group of people.

**32**